\documentclass[11pt]{article}
\usepackage[letterpaper, total={6.5in, 9in}]{geometry}
\usepackage{amsmath, amssymb}
\usepackage{bm}
\usepackage{commath}
\usepackage{graphicx}
\usepackage[tableposition=top]{caption}
\usepackage{subcaption}
\usepackage{keyfloat}
\usepackage[hidelinks]{hyperref}
\usepackage{doi}
\usepackage[ruled,vlined,linesnumbered]{algorithm2e}

\input{variables.sty}

\title{Improving gearshift controllers for electric vehicles with reinforcement learning}
\author{Marc-Antoine Beaudoin \thanks{Corresponding author: ma.beaudoin@mail.mcgill.ca} \thanks{Intelligent Automation Lab, Centre for Intelligent Machines, McGill University, Room 503, McConnell Engineering Building, 3480 University Street, Montreal, QC, Canada, H3A 0E9} \and Benoit Boulet \footnotemark[2]}
\date{December 1, 2021}

\begin{document}
\maketitle

\begin{abstract}
	During a multi-speed transmission development process, the final calibration of the gearshift controller parameters is usually performed on a physical test bench. Engineers typically treat the mapping from the controller parameters to the gearshift quality as a black-box, and use methods rooted in experimental design -- a purely statistical approach -- to infer the parameter combination that will maximize a chosen gearshift performance indicator. This approach unfortunately requires thousands of gearshift trials, ultimately discouraging the exploration of different control strategies. In this work, we calibrate the feedforward and feedback parameters of a gearshift controller using a model-based reinforcement learning algorithm adapted from \textsc{pilco}. Experimental results show that the method optimizes the controller parameters with few gearshift trials. This approach can accelerate the exploration of gearshift control strategies, which is especially important for the emerging technology of multi-speed transmissions for electric vehicles. \\
	
	\textbf{Keywords}: electric vehicle, multi-speed transmission, reinforcement learning, automatic tuning, gearshift controller
\end{abstract}

\section{Introduction}
To perform smooth and consistent gearshifts, a multi-speed transmission requires a well-designed and well-calibrated gearshift controller, whose development can be a challenge. For that, engineers first synthesize a controller: they define its mathematical structure and parametrization. Then, they find controller parameters that maximize a chosen performance objective. An initial set of parameters can be obtained from a principled method that rely on an approximate model of the transmission and vehicle dynamics. Often, the final set of controller parameters is calibrated from gearshift trials on a physical transmission test bench. At this point, engineers typically rely solely on statistics to infer the best combination of parameters from the recorded gearshift trials -- an approach centered around the design of experiments (DOE)~\cite{fisher_design_1935}. This approach can be time and resource consuming, sometimes requiring thousands of gearshift trials~\cite{mishra_automated_2021,kucukay_efficient_2009,boissinot_automated_2015}. This is because despite modern advances~\cite{jones_efficient_1998}, DOE-like methods treat the mapping from the controller parameters to the gearshift performance indicator as a black box. This leaves no choice but to generate a lot of data points by performing multiple gearshift trials with varied parameters, do statistical inference on the collected data set, and try optimize the parameters with this information. This leads to wonder whether modern approaches in reinforcement learning can be used to assist in the development of gearshift controllers by better leveraging the data generated during the gearshift trials. In particular, it should be possible to better exploit prior knowledge of the approximate system dynamics. \\

To be practically interesting, a learning approach to gearshift controller calibration should drastically reduce the number of gearshift trials required. Moreover, the method should yield controllers that perform well under varying operating conditions -- not just the specific conditions under which the training data was obtained. Because the number of gearshift trials would be reduced, engineers should be able to synthesize and iterate through a wide range of controller types with varying parametrization. This brings about a last requirement: the learning method should be easily amendable to various controller designs. This is important for the development of multi-speed transmissions for electric vehicles, as this emerging technology may not have converged to well established gearshift control strategies. In this article, we present a gearshift controller tuning method based on reinforcement learning, and argue that it is practically interesting for automotive engineers on the basis of the considerations introduced above.\\

For this research, we designed and parameterized a full-state linear feedback controller combined with a feedforward signal for the clutch-to-clutch gearshift of a multi-speed transmission for electric vehicles. The controller is detailed in Section~\ref{sec:controller}. We implemented the controller on a physical test bench and tuned its parameters using a variant of the \textsc{pilco} algorithm~\cite{deisenroth_pilco:_2011}, whose implementation details are presented in Section~\ref{sec:algorithm}. Experimental results are presented in Section~\ref{sec:results}. The rest of the introduction consists of motivating the choices of control strategy and learning algorithm for this study, as well as comparing the chosen approach to that of other research with similar objectives.

\subsection{The clutch-to-clutch gearshift control problem} \label{sec:gearshift_problem}
This article focuses on clutch-to-clutch gearshifts, and more precisely, power-on upshifts. Such a gearshift is exemplified in Figure~\ref{fig:nominal}. At the beginning of an upshift, the motor speed is synchronized with the gear 1 speed, Clutch 1 is engaged, and Clutch 2 is fully disengaged. At the end of the gearshift, the motor needs to be synchronized with the gear 2 speed, the torque on Clutch 1 reduced to zero, and Clutch 2 fully engaged.

Due to clutch nonlinearities, a clutch-to-clutch upshift must be done in two consecutive phases: the torque transfer phase, where the motor speed is maintained at or above the gear 1 synchronization speed, while Clutch 1 torque is gradually reduced to zero and Clutch 2 torque is increased; followed by the inertia phase, where the motor speed is reduced to match the gear 2 synchronization speed. We proved in~\cite{beaudoin_fundamental_2021} that in the context of electric vehicles, a deviation from this shifting strategy will cause unnecessary vehicle jerk.

If Clutch 1 is a one-way clutch, the motor speed will follow the gear~1 synchronization speed until the reaction torque on Clutch 1 is reduced to zero, at which point Clutch 1 automatically disengages, marking the end of the torque phase. If Clutch~1 is a friction clutch, it can be maintained in a sticking state during the torque phase, also resulting in the motor speed following the gear 1 synchronization speed. If this strategy is used, Clutch 1 should be released at the moment where its reaction torque is zero. This timing can be a challenge in practice due to the lack of direct measurement of the clutch torque. Alternatively, Clutch 1 can be made slipping during the torque phase. In this case, the motor speed should be maintained above the gear 1 synchronization speed, and Clutch 1 torque must be reduced to zero before the motor speed drops below the gear~1 synchronization speed~\cite{beaudoin_fundamental_2021}. In this article, Clutch 1 is a friction clutch, and we chose to have it slipping during the torque phase. However, the learning method introduced and the conclusions reached should also apply to the other strategies discussed above.\\

In the context of an electric vehicle, if Clutch 1 is a one-way clutch, there are two only actuators to control during the torque phase: the motor and Clutch~2. If Clutch 1 is a friction clutch, then it must also be controlled during the torque phase. In the inertia phase however, there are always only two actuators to control: the motor and Clutch 2. Every actuator can be controlled solely from either a feedforward or a feedback signal, or a combination of both. Oftentimes, two separate controllers are used for the torque and inertia phases. A typical choice is to use feedforward control for the torque phase, and to add a feedback component only during the inertia phase~\cite{bai_dynamic_2013}. In this work, we use the same controller for both phases, where Clutch 1 is controlled solely from a feedforward signal, and the motor and Clutch 2 are controlled from a combination of feedforward and feedback signals. This choice was made in order to demonstrate that the learning algorithm is capable of concurrently tuning the parameters of both the feedforward signals and the feedback controller. Again, the learning algorithm should also apply to the other cases discussed above.\\

In terms of feedback controller type and principled design method, several choices are reported in the literature. In~\cite{haj-fraj_optimal_2001}, researchers first linearized the system along the reference gearshift trajectory, then formulated an optimal control problem and used dynamic programming to solve it. The controller is a feedforward plus linear feedback controller. Researchers in~\cite{mousavi_seamless_2015} first obtained an open-loop optimal controller using the Pontryagin's Minimum Principle, and closed the feedback loop with the design of a backstepping controller. Article~\cite{gao_design_2011} presents a backstepping controller that integrates lookup tables for the strongly nonlinear elements of the powertrain model, such as the torque converter. In~\cite{gao_nonlinear_2011}, the same powertrain nonlinearities are considered for the design of a feedforward controller this time, which is used in combination with a linear feedback controller. In~\cite{sorniotti_analysis_2012}, the motor torque is controlled with a proportional-integral-derivative (PID) feedback controller during the inertia phase. The PID gains were tuned by shaping the closed-loop transfer function between the motor torque and the motor speed. In~\cite{sanada_study_1998}, researchers designed a robust feedforward-feedback controller for the inertia phase using $\mu$-synthesis. The solution is guaranteed robust stability and robust performance given the parametric uncertainty included in the model. Similarly, researchers in~\cite{kim_cooperative_2020} reported a complete gearshift solution which includes a multi-variable feedback controller designed with the robust $\mathcal{H}_\infty$ method. Finally, model predictive control was also used to solve the clutch-to-clutch gearshift problem~\cite{mesmer_embedded_2019}. In this work, the linear feedback controller used is initially tuned with the linear quadratic regulator (LQR) method, which is obtained from a nominal model of the linear system dynamics. This yields a feedback controller structure with an interesting number of parameters to tune -- the 8 entries of the $\Kc$ matrix, see Section~\ref{sec:controller}.\\

None of the methods reported above learn the feedback controller from iterated trials with a transmission test bench. The closest to it would be the work reported in~\cite{mishra_automated_2021}, where researchers use iterative learning control (ILC) to tune the parametrization of a feedforward signal for the closure of Clutch 2. The experimental results show that very few trials are required to learn appropriate parameters. However, ILC directly iterates on control signals~\cite{bristow_survey_2006}, therefore this method is ill-suited for tuning the feedback portion of gearshift controllers. The next section addresses this challenge. 

\subsection{Learning a gearshift controller}
This section addresses the problem of learning controller parameters from experience. The principled control design methods introduced in Section~\ref{sec:gearshift_problem} all rely on a system model to compute the controller parameters. It is customary to use system identification methods to obtain, calibrate, or validate system models~\cite{ljung_system_1999}. With an identified model, one can hope to get a superior controller since the design method now relies on a model that better represents the true system dynamics. In this work, we go beyond this approach and propose using reinforcement learning to concurrently and iteratively gain knowledge about the system dynamics and tune the control parameters. \\

Reinforcement learning and control engineering have essentially the same goal: to obtain an agent (a controller) that behaves optimally in a given environment by maximizing a reward (minimizing a cost function)~\cite{russell_artificial_2021}. Notably, central to both optimal control and reinforcement learning are Bellman's Principle of Optimality~\cite{bellman_dynamic_1957} and dynamic programming. Initially centered around problems with discrete state and action spaces, researchers in reinforcement learning now address continuous problems by leveraging the recent advances in function approximation -- the object of supervised learning, the main field of machine learning. 

Methods in reinforcement learning are typically classified in two categories: model-based and model-free approaches~\cite{sutton_reinforcement_2018}. In the former, the agent progressively builds an internal model of the environment (learning), then uses this model to design a control policy (planning). In the latter, the agent directly learns a control policy from interacting with the environment. Model-free approaches tend to require many more interactions with the environment~\cite{janner_when_2019}, hence our choice of a model-based approach for this work.\\

The learning algorithm we chose is \textsc{pilco}~\cite{deisenroth_pilco:_2011,deisenroth_efficient_2010}, which uses Gaussian processes (GP)~\cite{rasmussen_gaussian_2006,liu_gaussian_2018} to efficiently learn the system dynamics. To name a few, this method was used to efficiently tune linear controllers~\cite{deisenroth_learning_2011} and multivariate PID controllers~\cite{doerr_model-based_2017} for robotic arm applications. In \textsc{pilco}, the control policy is iterated with analytic gradients obtained from simulated policy rollouts using the learned model. However in this work, we make use of the automatic gradient functionalities of TensorFlow~\cite{abadi_tensorflow_2016} for additional speed and flexibility in the implementation of the method. The primary criticism of \textsc{pilco} is that the method scales poorly for problems of higher dimensions~\cite{wang_benchmarking_2019}, which should not be an issue in this work. 

For such problems of higher dimensions, an alternative would be guided policy search~\cite{levine_guided_2013}. In this method, control policies are randomly searched in a model-free fashion, but the search is guided by optimal control solutions obtained with differential dynamic programming~\cite{todorov_generalized_2005}, using a nominal model of the system dynamics. This method is interesting for avoiding local minima in complex high-dimensional control problems. But because our control problem is quite small, this method is likely to be less efficient than \textsc{pilco} and provide little added benefit. \\

Another alternative would be Coarse-ID control~\cite{dean_sample_2019}. This method starts with the identification of a linear system dynamics with least squares estimation. Then a bootstrap technique it proposed to bound the error between the real dynamics and the identified model. Finally, a controller is synthesized by solving a robust optimization problem. Researchers introduced a method for LQR controller synthesis, but the work could be extended to other controller types, and perhaps feedforward signals as well. Coarse-ID is very close to the traditional system identification plus principled controller synthesis method discussed at the beginning of this section, with the only difference being that machine learning is used twice: once for the system identification, and again for the uncertainty estimation. We subscribe to the idea that researchers should strive to reduce the gap between reinforcement learning and control theory~\cite{recht_tour_2019}. This study is an attempt to do so. 

\begin{figure}
	\centering
	\includegraphics[width=0.95\linewidth,trim={0cm 0.5cm 0cm 0.5cm},clip]{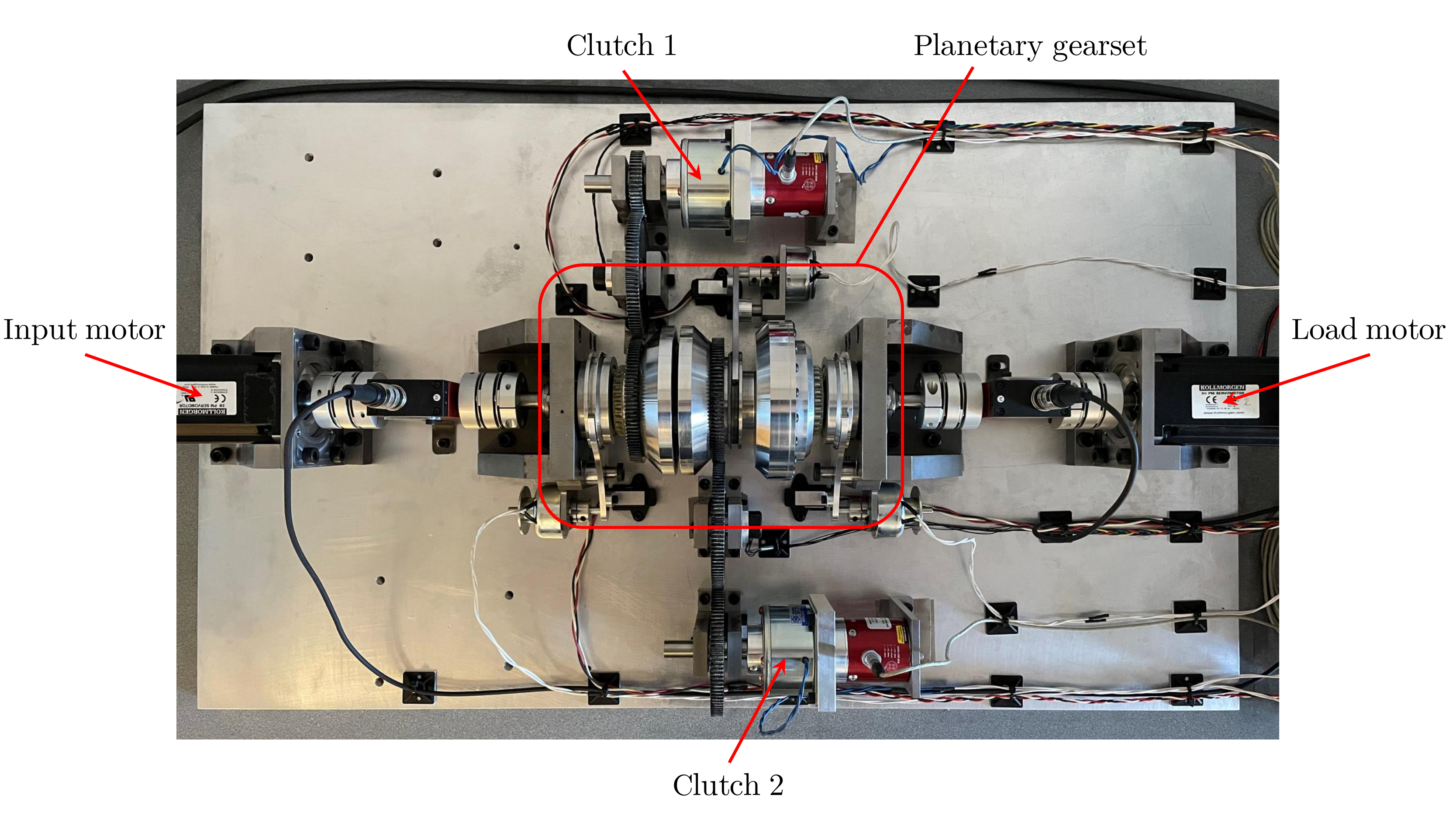}
	\caption{Reduced-scale electric vehicle transmission prototype.}
	\label{fig:experiment}
\end{figure}

\begin{figure}
	\centering
	\includegraphics[width=0.55\linewidth,trim={1.8cm 25.5cm 10.8cm 0.2cm},clip]{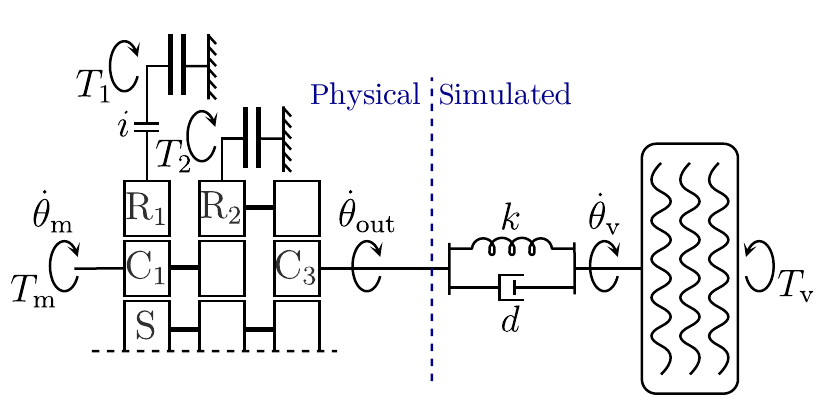}
	\caption{Driveline and vehicle model. The components on the left -- the motor, planetary gearset, and clutches -- are the physical components of Figure~\ref{fig:experiment}. The vehicle model on the right is simulated in real time, and used to generate the torque command on the load motor in Figure~\ref{fig:experiment}.}
	\label{fig:model}
\end{figure}

\section{Gearshift controller design and parametrization} \label{sec:controller}
The transmission test bench is presented on Figure~\ref{fig:experiment}, and the corresponding system model is shown on Figure~\ref{fig:model}. The system is in part physically realized with the input motor, the clutches, and the planetary gearset, and in part simulated in real-time with a simple driveline model. Both clutches are friction-plate electromagnetic brakes: a spring keeps the plates apart, until the electromagnet is activated, which magnetizes the floating plate and brings the braking surface into contact. The load motor on Figure~\ref{fig:experiment} inputs a torque that corresponds to the simulated driveline torque. In the displayed configuration, the planetary gearset is composed of the five rotating bodies labeled on Figure~\ref{fig:model}: S, $\mathrm{C}_1$, $\mathrm{C}_3$, $\mathrm{R}_1$, $\mathrm{R}_2$. The torque inputs on the system are the motor torque $\Tm$, Clutch~1 torque $T_1$, Clutch 2 torque $T_2$, and a resistive vehicle torque $\Tv$. Choosing the motor speed $\tpm$, the output shaft speed $\tpout$ and the vehicle speed $\tpv$ as the general coordinates, the equations of motion are
\begin{align}
	\tppm &= c_1 \Tm + c_2\big(k(\tout-\tv) + d(\tpout-\tpv)\big) + c_3 T_1 + c_4 T_2,\\
	\tppout &= c_5 \Tm + c_6\big(k(\tout-\tv) + d(\tpout-\tpv)\big) + c_7 T_1 + c_8 T_2, \\
	\tppv &= \Iv^{-1} \big(k(\tout-\tv) + d(\tpout-\tpv)\big) - \Iv^{-1} \Tv,
\end{align}
where $c_1$ to $c_8$ are constants that regroup parameters such as the inertia of the rotating elements and the number of teeth on the meshing ones. These constants are found analytically during the routine process of obtaining the equations of motion, such as using the Newton-Euler approach for instance. The parameter $\Iv$ represents an equivalent vehicle inertia projected on a rotating body, assuming no wheel slip. The parameter $k$ is the equivalent driveline stiffness and $d$, its damping. By choosing the set of states $\bm{\mathrm{x}} = [\tpm, \tpout, \tpv, (\tout-\tv)]^\top$ and control inputs $\bm{\mathrm{u}} = [\Tm, T_1, T_2]^\top$ for the system, a linear state space representation of its dynamics can be obtained:
\begin{align}
	\dot{\bm{\mathrm{x}}} &= A \bm{\mathrm{x}} + B \bm{\mathrm{u}} + \tzero, \label{eq:ss_1}\\[3pt]
	A &= \begin{bmatrix}
		0 & c_2 d & -c_2 d & c_2 k \\
		0 & c_6 d & -c_6 d & c_6 k \\
		0 & \Iv^{-1}d & - \Iv^{-1}d & \Iv^{-1}k \\
		0 & 1 & -1 & 0 
	\end{bmatrix}, \quad
	B = \begin{bmatrix}
		c_1 & c_3 & c_4 \\
		c_5 & c_7 & c_8 \\
		0 & 0 & 0 \\
		0 & 0 & 0
	\end{bmatrix}, \\
	\tzero &= [0, 0, -\Iv^{-1} \Tv, 0]^\top \label{eq:ss_3}.
\end{align} 

Since they are simulated, the parameters $\Iv$, $k$, and $d$ must be defined. First, $\Iv$ is set such that, in first gear, the equivalent inertia of the motor and transmission projected at the vehicle level is 10\% that of $\Iv$ -- a realistic ratio for a real-world vehicle~\cite{sorniotti_analysis_2012,galvagno_drivability_2013,holdstock_linear_2013}. Then, $k$ and $d$ are set such that, in first gear again, the natural frequency of the driveline is 5\,Hz, and its damping ratio is 0.15, which are also typical values of driveline dynamics. \\

\begin{figure}
	\begin{subfigure}[t]{.5\textwidth}
		\centering
		\includegraphics[width=0.9\linewidth,trim={0cm 0cm 0cm 0cm},clip]{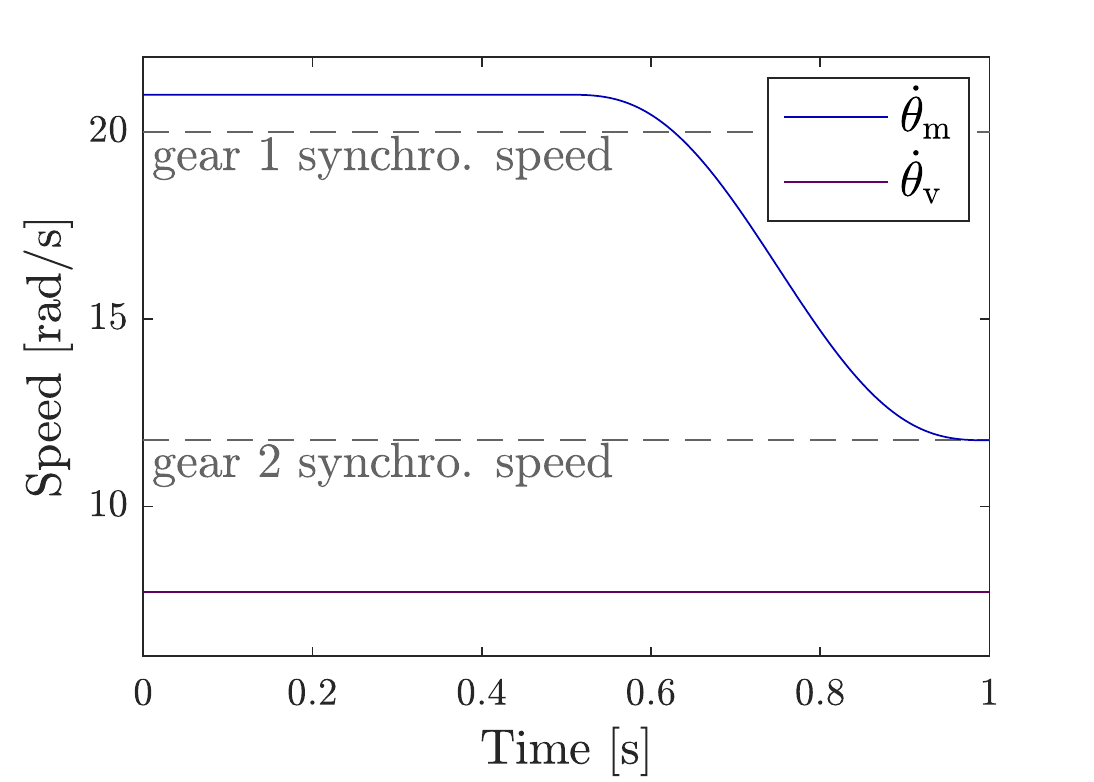}
		\caption{Prescribed trajectory $\bar{\bx}$}
		\label{fig:trajectory}
	\end{subfigure}
	\begin{subfigure}[t]{.5\textwidth}
		\centering
		\includegraphics[width=0.9\linewidth,trim={0cm 0cm 0cm 0cm},clip]{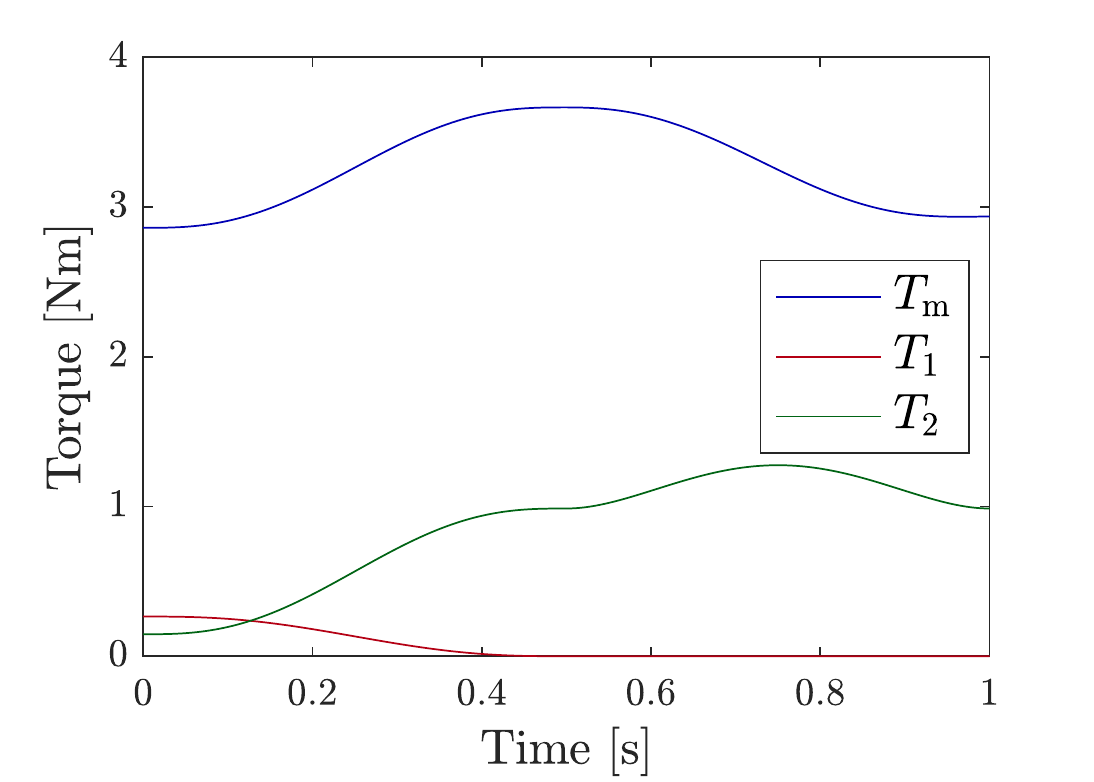}
		\caption{Nominal torque command $\bar{\bu}$}
		\label{fig:command}
	\end{subfigure}
	\caption{The gearshift begins with a torque phase $[0\, \mathrm{s} - 0.5\, \mathrm{s}]$, where $\tpm$ is kept above the gear 1 synchronization speed and $T_1$ is gradually reduced to zero. It ends with an inertia phase $[0.5\, \mathrm{s} - 1\, \mathrm{s}]$, where $\tpm$ is brought down to the gear 2 synchronization speed.}
	\label{fig:nominal}
\end{figure}

A prescribed gearshift trajectory $\bar{\bx}$ and a nominal torque command $\bar{\bu}$ can be computed from the Equations~\eqref{eq:ss_1}-\eqref{eq:ss_3}; they are shown on Figure~\ref{fig:nominal}. The controller's objective is to track the prescribed state trajectory $\bar{\bx}$. In this study, the main gearshift performance indicator is maintaining a constant vehicle speed $\tpv$. The gearshift begins under the following conditions: $\tpm$~=~20\,rad/s and $\Tv$~=~3.5\,Nm. For this study, $\Tv$ is kept constant throughout the gearshift, which is a common assumption in gearshift control research~\cite{mousavi_seamless_2015,kim_cooperative_2020}. The motor speed $\tpm$ is kept 1\,rad/s above the gear~1 synchronization speed during the torque phase, then is smoothly brought down during the inertia phase. A constant output shaft speed $\tpout$ that matches $\tpv$ is prescribed, as well as a constant driveshaft elongation $(\tout-\tv) = \Tv / k$. With $\bar{\bx}$ defined, the next step is to compute $\bar{\bu}$. First, an ``idealized" nominal torque command defined as $\bar{\bm{\mathrm{u}}}_0 = [T_{\mathrm{m},0}, T_{1,0}, T_{2,0}]^\top$ is computed directly form Equations~\eqref{eq:ss_1}-\eqref{eq:ss_3}. For that, an arbitrary trajectory is imposed for $T_{1,0}$ -- it starts at the clutch 1 torque at the beginning of the gearshift, and ends at zero at the end of the torque phase. Then $T_{\mathrm{m},0}$ and $T_{2,0}$ can be computed by solving for the remaining terms in Equations~\eqref{eq:ss_1}-\eqref{eq:ss_3}. Finally, the ``actual" $\bar{\bu}$ is obtained as follows:
\begin{align}
	T_{\mathrm{m}} & = T_{\mathrm{m},0} + a_1\tpm/\tpout + a_2, \\
	T_{1} & = a_3 T_{1,0}, \\
	T_{2} & = a_4 T_{2,0},
\end{align}
where $\{a_1,\ldots,a_4\}$ are parameters for the feedforward signal $\bar{\bu}$ of the gearshift controller. These parameters help to account for missing terms in Equations~\eqref{eq:ss_1}-\eqref{eq:ss_3} such as friction in the planetary gearset, as well as other discrepancies between the nominal model and the real system dynamics. These are the four parameters the learning algorithm will vary in order to tune $\bar{\bu}$. An initial value for these parameters is defined heuristically from simple measurements done on the test bench, such as estimating friction from constant-speed runs. \\

\begin{figure}
	\centering
	\includegraphics[width=0.50\linewidth,trim={2.5cm 23.2cm 9.2cm 2.2cm},clip]{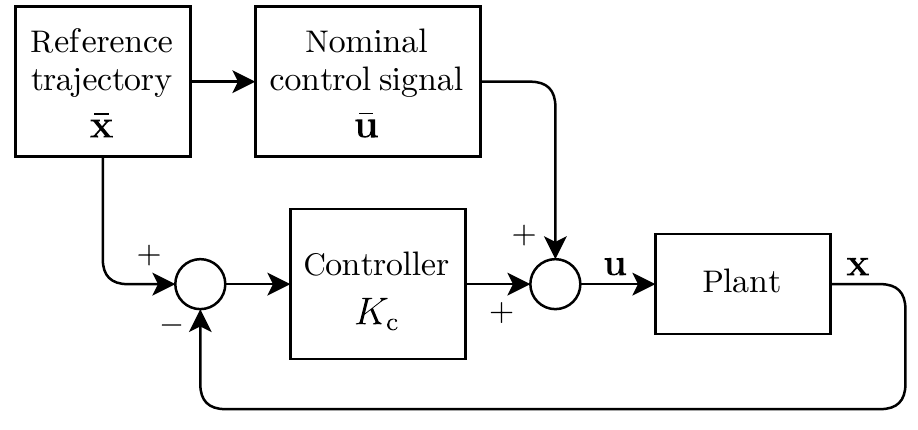}
	\caption{Gearshift controller setup. The control signal $\bu = \bar{\bu} + \Kc(\bar{\bx} - \bx)$ is the sum of a feedforward $\bar{\bu}$ and a state feedback components. }
	\label{fig:control}
\end{figure}

The complete controller has the form
\begin{equation}
	\bu = \bar{\bu} + \Kc(\bar{\bx} - \bx), \label{eq:control}
\end{equation} 
where $\bu$ is the control signal, $\bar{\bu}$ is the (feedforward) nominal control signal, and $\Kc(\bar{\bx} - \bx)$ is the feedback term. The control loop is pictured on Figure~\ref{fig:control}. The linear controller $\Kc$ is obtained by solving an LQR problem where the second column of $B$ is removed -- recall that only $\Tm$ and $T_2$ are feedback controlled. Even when removing the second column of $B$, the system is still controllable. The resulting controller is a $2\times4$ matrix, which adds eight more controller parameters to tune during training. The next section presents the algorithm used to tune the 12 controller parameters $\bpsi$ from gearshift trials.

\section{Learning algorithm} \label{sec:algorithm}
\begin{figure}
	\centering
	\includegraphics[width=0.65\linewidth,trim={0.7cm 20.4cm 9.2cm 2.6cm},clip]{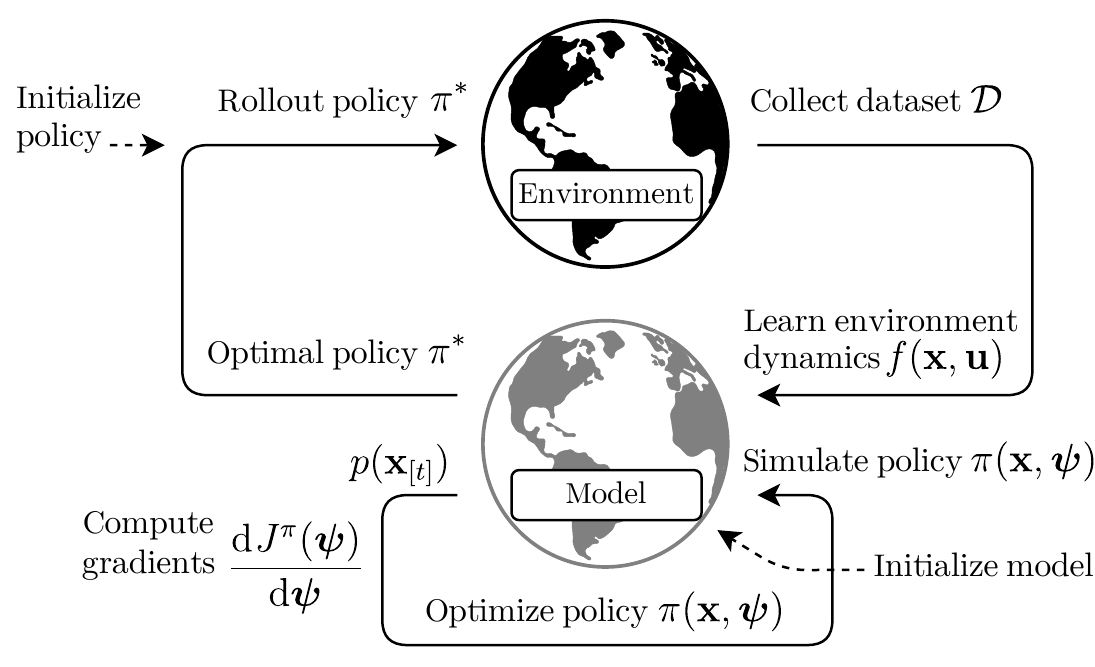}
	\caption{Schematic representation of the model-based reinforcement learning algorithm used to tune the parameters $\bpsi$ of the control policy $\pi$.}
	\label{fig:rlproblem}
\end{figure}

The learning algorithm used in this work is an altered version of \textsc{pilco}, schematically represented in Figure~\ref{fig:rlproblem}, and outlined in Algorithm~\ref{alg:pilco}. The learning problem is formulated in discrete time; the states and control actions are still continuous. The system model used for the simulated policy rollouts is
\begin{align}
	\xt &=  \Ad \xtm + \Bd \utm + f(\xtm,\utm) + \tzero, \label{eq:sys}\\
	\utm &= \pi(\xtm,\bpsi) = \ubtm + \Kc (\xbtm - \xtm), \label{eq:ut}
\end{align}
where $\Ad$, $\Bd$, and $\tzero$ are obtained by discretizing the system in Equations~\eqref{eq:ss_1}-\eqref{eq:ss_3}. The dynamics of Equation~\eqref{eq:sys} is composed of a known nominal model ($\Ad \xtm + \Bd \utm + \tzero$) and an unknown function $f(\xtm,\utm)$. This unknown dynamics $f(\bx,\bu)$ is to be learned from gearshift trials on the test bench, which is addressed in Section~\ref{sec:GP}. The control policy $\pi(\bx,\bpsi)$ is deterministic, and~$\bpsi$ regroups the 12 policy parameters. The tuning of these parameters is the subject of Section~\ref{sec:cost}.

\begin{algorithm}
	\caption{\textsc{pilco} for gearshift controllers} \label{alg:pilco}
	\KwResult{Learned policy $\pi(\bx,\bpsi)$.}
	Initialize $\bpsi$: initialize $\Kc$ from LQR with a nominal model, initialize $\bar{\bu}$ heuristically.\\
	\While{$\pi$ not learned}{
		Rollout the policy $\pi$ on the test bench and collect a dataset $\mathcal{D}$.\\
		Learn the unknown system dynamics $f(\bx,\bu)$ with Gaussian processes.\\
		\While{$\pi$ not optimized}{
			Simulate a policy rollout: compute the state probability distribution $p(\xt) \, \forall t\in\{0,\ldots,T\}$, the cost function $J^\pi(\bpsi)$, and the gradients $\od{J^\pi(\bpsi)}{\bpsi}$.\\
			Iterate the policy parameters $\bpsi$ using the gradients $\od{J^\pi(\bpsi)}{\bpsi}$.
		}
	}
\end{algorithm}

\subsection{Gaussian process regression}\label{sec:GP}
The unknown dynamics $f(\bx,\bu)$ is learned using Gaussian processes. GPs approximate functions with a scalar output, thus $D$ functions $f_d(\bx,\bu)$ are learned, i.e., one for every state we wish to predict. This means $f(\bx,\bu) = [f_1(\bx,\bu), \ldots, f_D(\bx,\bu)]^\top$. For each of the dimensions $d$, $n$ training targets and feature vectors are obtained with
\begin{align}
	y_{d\,[i]} &= x_{d\,[i]} - [\Ad \xim + \Bd \uim + \tzero]_d, \\
	\bz_{[i]} &= [\xim^\top, \uim^\top]^\top.
\end{align}
Each dimension $d$ has its own target vector $\by_d$ composed of the $y_{d\,[i]}$ elements computed above. All dimensions share the same set of corresponding feature vector $Z = [\bzone,\ldots,\bzn]$. Effectively, the GP learns the difference between the nominal model and the real (measured) system dynamics. GPs are stochastic processes characterized by a mean  $m(\bz)$ and a kernel function $k(\bz,\bz')$.  Here, we choose the mean function $m(\bz) := 0$, and the square exponential kernel function
\begin{equation}
	k(\bz,\bz') := \sigmaf^2 \exp(-\tfrac{1}{2}(\bz - \bz')^\top \Lambda^{-1}(\bz - \bz')),
\end{equation}
where $\sigmaf^2$ is the signal variance and $\Lambda = \mathrm{diag}([l_1^2,\ldots,l_{D+F}^2])$ is a diagonal matrix composed of the characteristics length-scales. These are the hyper-parameters of the Gaussian process, and each dimension $d$ has its own set of hyper-parameters. The last hyper-parameter in this problem is the noise variance $\sigma_\epsilon^2$, which will appear in Equations~\eqref{eq:mu}~and~\eqref{eq:sig}. \\

In this work, the system dynamics is modeled with a linear nominal model and a Gaussian process with a zero mean function $m(\bz) = 0$. This is equivalent to modeling the dynamics with no nominal model and a Gaussian process with a linear mean function. In the original implementation of \textsc{pilco}~\cite{deisenroth_pilco:_2011}, no nominal model is used and the mean function is also zero. This means that the entirety of the system dynamics has to be learned from data. Unsurprisingly, researchers in~\cite{bischoff_policy_2014} showed that using a linear model as a mean function accelerates the learning process, which motivates the use of the linear nominal model in our case. \\

For a deterministic test point $\bzs$, the output of $f_d(\bz)$ will be normally distributed with mean and variance

\begin{align}
	\mu_d(\bz_*) &= \Kzsz(\Kzz + \sigma_\epsilon^2 I)^{-1} \by_d, \label{eq:mu}\\
	\Sigma_d(\bz_*) &= k(\bz_*,\bz_*) - \Kzsz(\Kzz + \sigma_\epsilon^2 I)^{-1} \Kzsz^\top, \label{eq:sig} \\
	\mathrm{where} \quad \Kzz &= \begin{bmatrix}
		k(\bzone,\bzone) & \cdots & k(\bzone,\bzn) \\
		\vdots & \ddots & \vdots \\
		k(\bzn,\bzone) & \cdots & k(\bzn,\bzn)
	\end{bmatrix}, \\
	\Kzsz &= \begin{bmatrix}
		k(\bz_*,\bzone) & \cdots & k(\bz_*,\bzn)
	\end{bmatrix}.
\end{align}

The accuracy of the prediction $f_d(\bz_*)$ depends on having appropriate GP hyper-parameters. It is possible to tune the set of hyper-parameters $\btheta_d$ by maximizing the logarithm of the marginal likelihood of the observed data points $\by_d$, such as suggested in~\cite{rasmussen_gaussian_2006}, with
\begin{equation}
	\log p(\by_d \vert Z,\btheta_d) = -\frac{1}{2} \by_d^\top K^{-1} \by_d - \frac{1}{2} \log \vert K \vert - \frac{n}{2} \log 2 \pi,
\end{equation}
where $K = \Kzz + \sigma_\epsilon^2 I$. The likelihood is maximized with a gradient-based optimization algorithm using 
\begin{align}
	\dpd{}{\theta_{d\,[j]}} \log p(\by_d \vert Z,\btheta_d) &= \frac{1}{2} \tr \bigg( \big( \balpha \balpha^\top - K^{-1}\big) \dpd{K}{\theta_{d\,[j]}} \bigg), \\[2pt]
	\balpha &= K^{-1} \by_d.
\end{align}

\subsection{Simulated policy rollouts and uncertainty propagation}
The next step in Algorithm~\ref{alg:pilco} is to simulate a rollout from an initial state $\bx_0$, under a given control policy $\pi$, from time $t=0$ to $t=T$. This means computing the state distributions $\{p(\bx_{[0]}), \ldots, p(\bx_{[T]})\}$. Computing $p(\xt)$ by performing the integration
\begin{equation}
	p(\xt) = \iint p(\xt|\xtm,\utm) p(\utm|\xtm) p(\xtm) \dif \xtm \dif \utm
\end{equation}
is generally intractable. In particular, $p(\xt|\xtm,\utm)$ is quite challenging. Recall that the results of Equations~\eqref{eq:mu}~and~\eqref{eq:sig} were for a deterministic test input $\bzs$. When $\xtm$ and $\utm$ are non-deterministic, the output of a Gaussian process is in general not Gaussian, even if $\xtm$ and $\utm$ are themselves normally distributed. In the original \textsc{pilco} implementation, they approximate the output distribution as Gaussian, and they obtain the mean and variance of the distribution using exact moment matching~\cite{deisenroth_pilco:_2011}. In this work, we take a somewhat simpler approach: we also approximate the output as a Gaussian distribution, but we get the mean and variance using an approximate solution. Assuming $\bzs \sim \mathcal{N}(\muzs,\Szs)$, and using a first order Taylor expansion around~$\muzs$, see~\cite{girard_gaussian_2002}, we have 
\begin{align}
	\mu_d(\bzs) &= \mu_d(\muzs), \\
	\Sigma_d(\bzs) &= \Sigma_d(\muzs) + \eval{\dpd{\mu_d(\bzs)}{\bzs}}_{\bzs = \muzs} \Szs \eval{\dpd{\mu_d(\bzs)}{\bzs}}_{\bzs = \muzs}^\top.	
\end{align}
This approach is also chosen by researchers in~\cite{hewing_cautious_2020}, where GPs are used in the context of model predictive control. This requires us to introduce the derivative of the mean of the GP prediction with respect to a deterministic test point $\bz_*$
\begin{equation}
	\dpd{\mu_d(\bzs)}{\bzs} = -(\Kzsz \odot \by_d^\top(\Kzz + \sigma_\epsilon^2 I)^{-1}) \tilde{Z}_*^\top \Lambda^{-1}, \label{eq:delmud}
\end{equation}
where $\tilde{Z}_* = [\bz_*-\bzone, \ldots, \bz_*-\bzn]$, and $\odot$ represents an element-wise product. \\

Assuming that $\xtm \sim \mathcal{N}(\muxtm,\Sxtm)$, the mean and variance of the control signals are
\begin{align}
	\muutm &= \ubtm + \Kc (\xbtm - \muxtm), \\
	\Sutm &= \Kc \Sxtm \Kc^\top.
\end{align}
Then $p(\ztm)$ can be expressed as
\begin{equation}
	\ztm \sim \mathcal{N} \left(
	\begin{bmatrix}
		\muxtm \\
		\ubtm + \Kc (\xbtm - \muxtm)
	\end{bmatrix},
	\begin{bmatrix}
		\Sxtm & \Sxtm \Kc^\top \\
		\Kc \Sxtm & \Kc \Sxtm \Kc^\top
	\end{bmatrix} \right).
\end{equation}
For convenience, Equations~\eqref{eq:sys}~and~\eqref{eq:ut} are regrouped into
\begin{equation}
	\xt = (\Ad - \Bd \Kc)\xtm + f(\xtm,\utm) + \Bd(\ubtm + \Kc \xbtm) + \tzero,
\end{equation}
from which an expression for $p(\xt)$ can be obtained as
\begin{align}
	\muxt &= (\Ad - \Bd \Kc)\muxtm + \muf(\muztm) + \Bd(\ubtm + \Kc \xbtm) + \tzero, \label{eq:trmu}\\
	\Sxt &= (\Ad - \Bd \Kc) \Sxtm (\Ad - \Bd \Kc)^\top + \Sigmaf(\ztm), \label{eq:trs}
\end{align}
where 
\begin{align}
	\muf(\muztm) &= [\mu_1(\muztm), \ldots, \mu_D(\muztm)]^\top, \\
	\Sigmaf(\ztm) &= \mathrm{diag}([\Sigma_1(\ztm),\ldots,\Sigma_D(\ztm)]).
\end{align}

\subsection{Cost function and its gradients}\label{sec:cost}
As is customary in reinforcement learning, the goal of Algorithm~\ref{alg:pilco} is to minimize the expected long-term cost of following a policy $\pi$ over a finite horizon of $T$ time steps 
\begin{equation}
	J^\pi(\bpsi) = \sum_{t=0}^{T}\mathbb{E}_{\xt}[c(\xt)].
\end{equation}
Following \textsc{pilco}'s original paper, the cost function used in this study is the saturating immediate cost 
\begin{equation}
	c(\xt) = 1- \exp \big( -\tfrac{1}{2}(\xt - \xbt)^\top L^{-1} (\xt - \xbt) \big),
\end{equation}
where $L^{-1}$ is a diagonal matrix whose elements dictate the width of the cost function for each of the state dimensions. For $\xt \sim \mathcal{N}(\muxt,\Sxt)$, the expectation of this cost function is
\begin{align}
	\mathbb{E}_{\xt}[c(\xt)] &= 1 - \vert I + \Sxt L^{-1} \vert^{-1/2} \exp \big( -\tfrac{1}{2}(\muxt - \xbt)^\top \tilde{S} (\muxt - \xbt) \big), \\[2pt]
	\tilde{S} &= L^{-1} (I + \Sxt L^{-1})^{-1}.
\end{align}

The cost can be minimized by following the gradients given by
\begin{equation}
	\dod{J^\pi(\bpsi)}{\bpsi} = \sum_{t=0}^{T} \dod{}{\bpsi} \mathbb{E}_{\xt}[c(\xt)]. \label{eq:grad}
\end{equation}
The two most viable options for computing these gradients are analytical differentiation and automatic differentiation. The original implementation of \textsc{pilco} computes the gradients analytically, which consists of expanding Equation~\eqref{eq:grad} with the chain rule until we obtain analytical expressions that can be computed directly. The first expansion is
\begin{equation}
	\dod{}{\bpsi} \mathbb{E}_{\xt}[c(\xt)] = \left( \dpd{}{{\muxt}} \mathbb{E}_{\xt}[c(\xt)] \right) \dod{{\muxt}}{\bpsi} + 
	\left( \dpd{}{{\Sxt}} \mathbb{E}_{\xt}[c(\xt)] \right) \dod{{\Sxt}}{\bpsi},
\end{equation}
where the derivatives of the mean and variance of the state distribution with respect to the controller parameters can be further expanded as
\begin{align}
	\dod{{\muxt}}{\bpsi} &= \dpd{{\muxt}}{{\muxtm}} \dod{{\muxtm}}{\bpsi} 
	+ \dpd{{\muxt}}{{\Sxtm}} \dod{{\Sxtm}}{\bpsi} + \dpd{{\muxt}}{\bpsi}, \label{eq:dmudpsi}\\[2pt]
	\dod{{\Sxt}}{\bpsi} &= \dpd{{\Sxt}}{{\muxtm}} \dod{{\muxtm}}{\bpsi} 
	+ \dpd{{\Sxt}}{{\Sxtm}} \dod{{\Sxtm}}{\bpsi} + \dpd{{\Sxt}}{\bpsi}. \label{eq:dsdpsi}
\end{align}
We notice that $\od{{\muxtm}}{\bpsi}$ and $\od{{\Sxtm}}{\bpsi}$ are given from the previous time step, while the rest of the gradients have to be further expanded. This approach quickly becomes cumbersome, and more importantly, several steps would need to be redone if the parametrization of the controller changed. 

For that reason, we implemented automatic differentiation~\cite{baydin_automatic_2018}. This method uses the fact that all computations are ultimately compositions of elementary operations with known derivatives. Automatic differentiation consists of augmenting the computation of the elementary operations leading to a result (here, the cost function) with the computation of the derivative of these operations. Then the stored derivatives can be combined with the chain rule to yield the derivative of the result with respect any chosen constituent of the computation. Several frameworks exists for implementing automatic differentiation; here we chose TensorFlow. In TensorFlow, computational graphs are used to perform the forward computations and the automatic differentiation efficiently. \\

In this study, we still worked out the analytical solution for the derivative of the cost function with respect to the matrix $\Kc$. The results are presented in Appendix~\ref{apx:grad}. We used the analytical solution to verify our implementation of automatic differentiation -- programming mistakes are easy to make and hard to detect otherwise. The results of Appendix~\ref{apx:grad} can be reused by the interested researchers to verify their own implementation of \textsc{pilco} with automatic gradients. Alternatively, numerical gradients can also be used for gradient validation. However, numerical gradients are never exact, and it can be hard to decipher whether the discrepancies are caused by numerical imprecision or programming errors. 

\section{Experimental results} \label{sec:results}
\begin{figure}
	\begin{subfigure}[t]{.5\textwidth}
		\centering
		\includegraphics[width=1\linewidth,trim={0cm 0cm 0cm 0cm},clip]{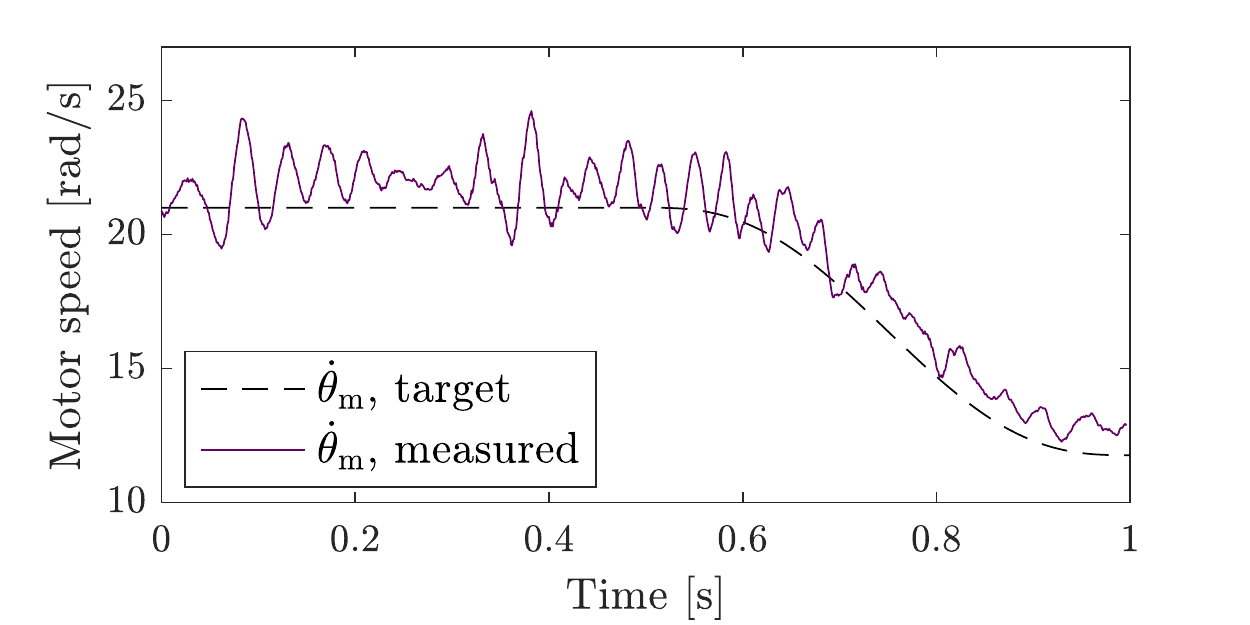}
		\caption{Measured motor speed $\tpm$}
		\label{fig:motor}
	\end{subfigure}
	\begin{subfigure}[t]{.5\textwidth}
		\centering
		\includegraphics[width=1\linewidth,trim={0cm 0cm 0cm 0cm},clip]{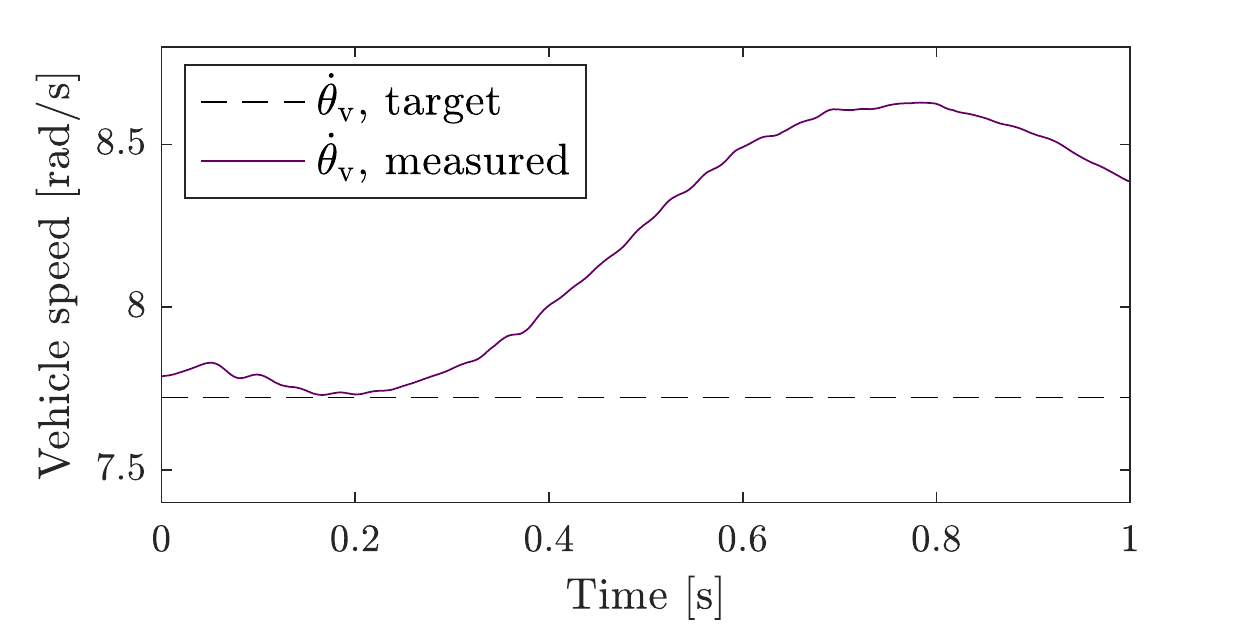}
		\caption{Measured vehicle speed $\tpv$}
		\label{fig:vehicle}
	\end{subfigure}
	\caption{Experimental results for the untrained control policy.}
	\label{fig:results}
\end{figure}

Measurements of $\tpm$ and $\tpv$ during a gearshift with the initialized (untrained) controller is shown on Figure~\ref{fig:results}. The rest of the result section focuses on the improvement of the tracking performance for~$\tpv$ with the proposed learning method. Figure~\ref{fig:iterations} shows the evolution of the trajectories of~$\tpv$ through the iterations of the bigger loop in Figure~\ref{fig:rlproblem} and Algorithm~\ref{alg:pilco}. Every trace is a measurement of~$\tpv$ on the test bench where the gearshift is performed with a newly optimized policy~$\pi^*$. Table~\ref{tbl:error} shows the reduction of various norms of the tracking error signal $e(t)$ for $\tpv$. The results show that very few gearshift trials (in this case, only about four) are required to tune the 12 parameters of the gearshift controller. The computation of each iterated policy~$\pi^*$ only takes about $100 \, \mathrm{s}$ on a laptop computer. Of course, the various measures of error reduction presented in Table~\ref{tbl:error} heavily depend on the quality of the initialized controller. After all, \textsc{pilco} was shown to be capable of learning controllers starting from randomly initialized parameters. In the context of a gearshift controller development process however, it may be counterproductive to randomly initialize the controllers given that several principled design methods exist in the literature, and engineers typically have good approximate models for the driveline dynamics. Therefore, it is interesting to see that the method still improves the performance of a reasonably initialized gearshift controller, and does so using only a few gearshift trials. \\

\begin{keyfloats}{2}
	\begin{keyfigure}{w=0.42\linewidth, c=\centering Evolution of vehicle speed $\tpv$ trajectories through the iterations., l=fig:iterations}
		\includegraphics[width=1\linewidth,trim={0.0cm 0cm 0.0cm 0.5cm},clip]{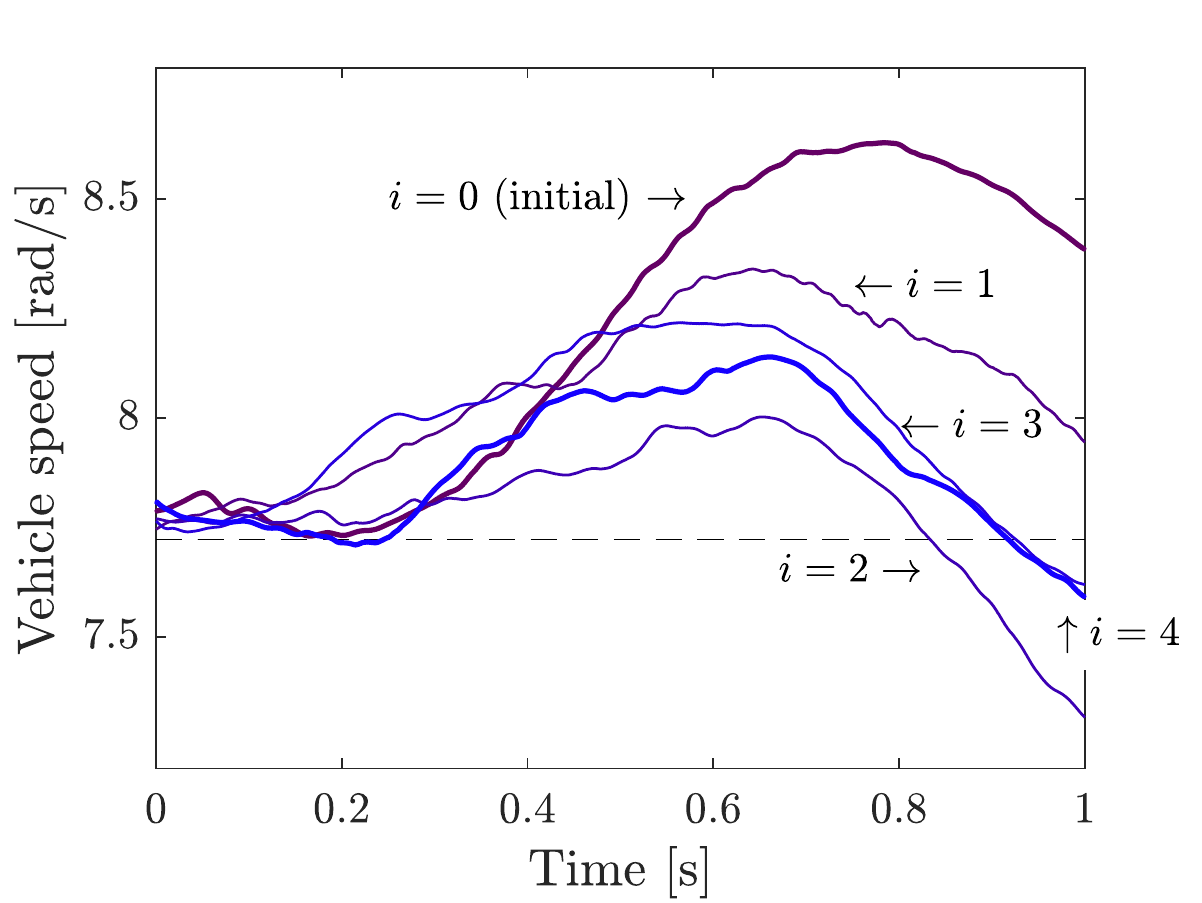}\label{fig:iterations}
	\end{keyfigure}
	\keytab{c=\centering Reduction in the tracking error of $\tpv$ through the iterations., l=tab:error}{
		\begin{tabular}{c|c|c|c} \label{tbl:error}
			iter. nb. ($i$) & $\Vert e \Vert_\infty $ & 
			$\vert e(1) \vert $ & $\Vert e \Vert_2 $ \\ \hline
			0& 0.91  &  0.66  & 18.5 \\
			1& 0.62  &  0.22  & 12.4 \\
			2& 0.41  &  0.41  &  5.4 \\
			3& 0.50  &  0.10  & 10.0 \\
			4& 0.42  &  0.13  &  7.3 \\ \hline
			reduction & 54 \% & 80 \% & 61 \% \\ \hline
	\end{tabular}}
\end{keyfloats}

\begin{figure}
	\begin{subfigure}[t]{.33\textwidth}
		\centering
		\includegraphics[width=1\linewidth,trim={0cm 0cm 0.5cm 0cm},clip]{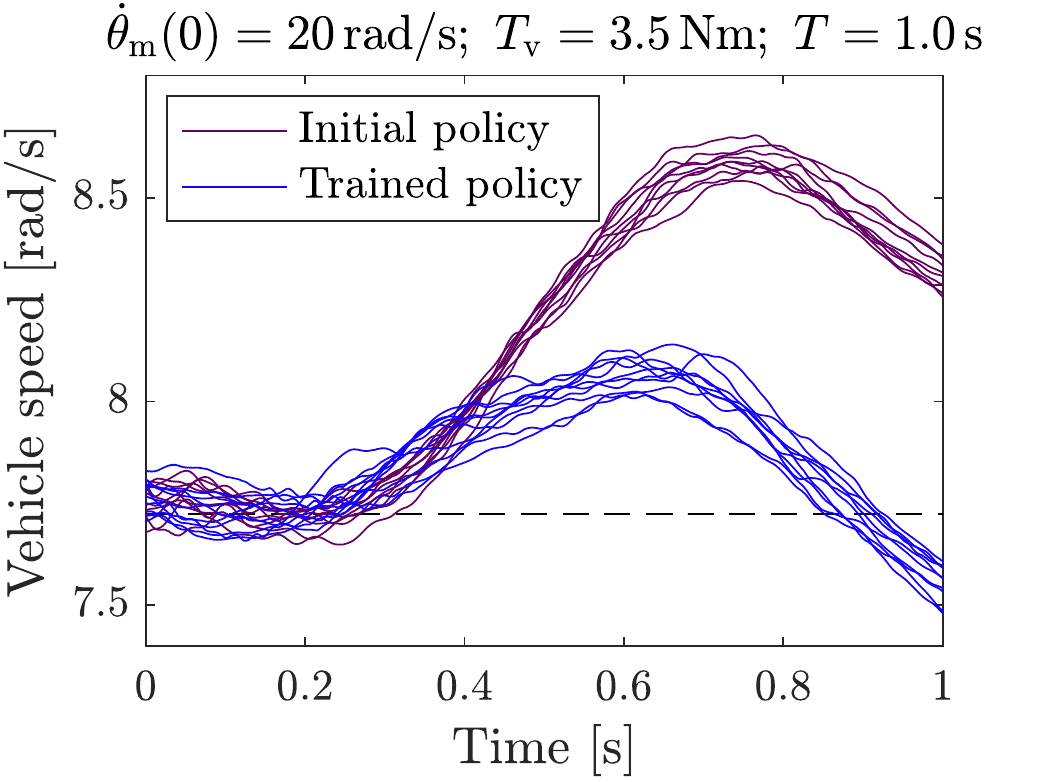}
		\caption{}
		\label{fig:shift_1}
	\end{subfigure}
	\begin{subfigure}[t]{.33\textwidth}
		\centering
		\includegraphics[width=1\linewidth,trim={0cm 0cm 0.5cm 0cm},clip]{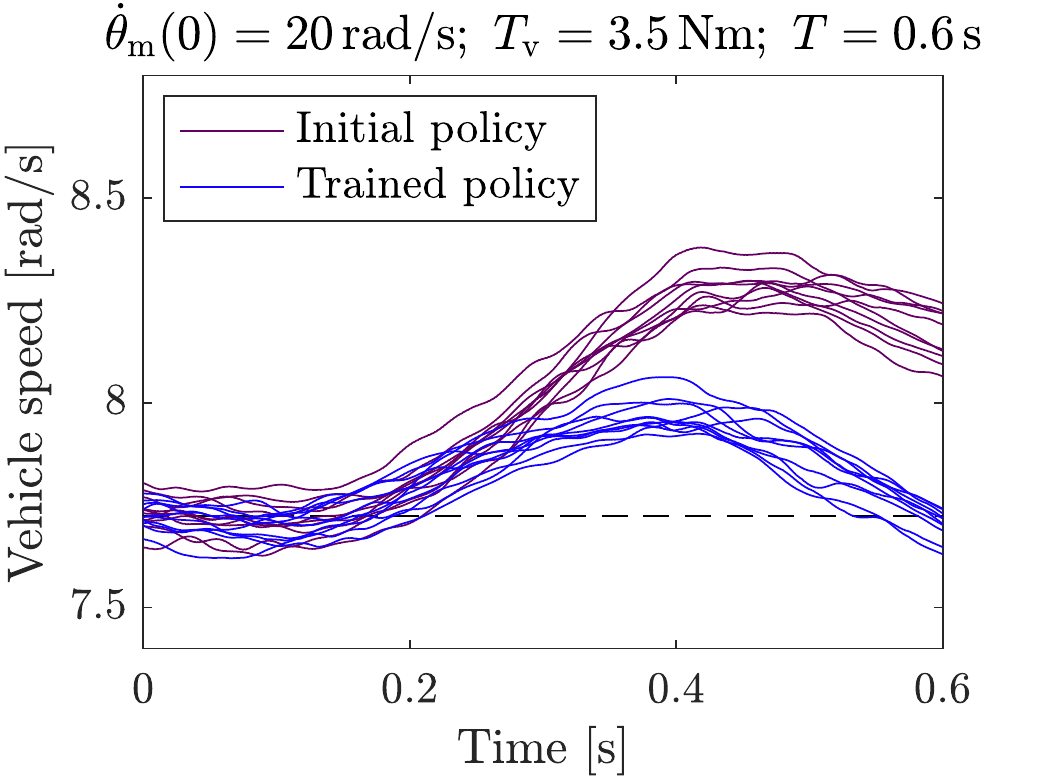}
		\caption{}
		\label{fig:shift_2}
	\end{subfigure}
	\begin{subfigure}[t]{.33\textwidth}
		\centering
		\includegraphics[width=1\linewidth,trim={0cm 0cm 0.5cm 0cm},clip]{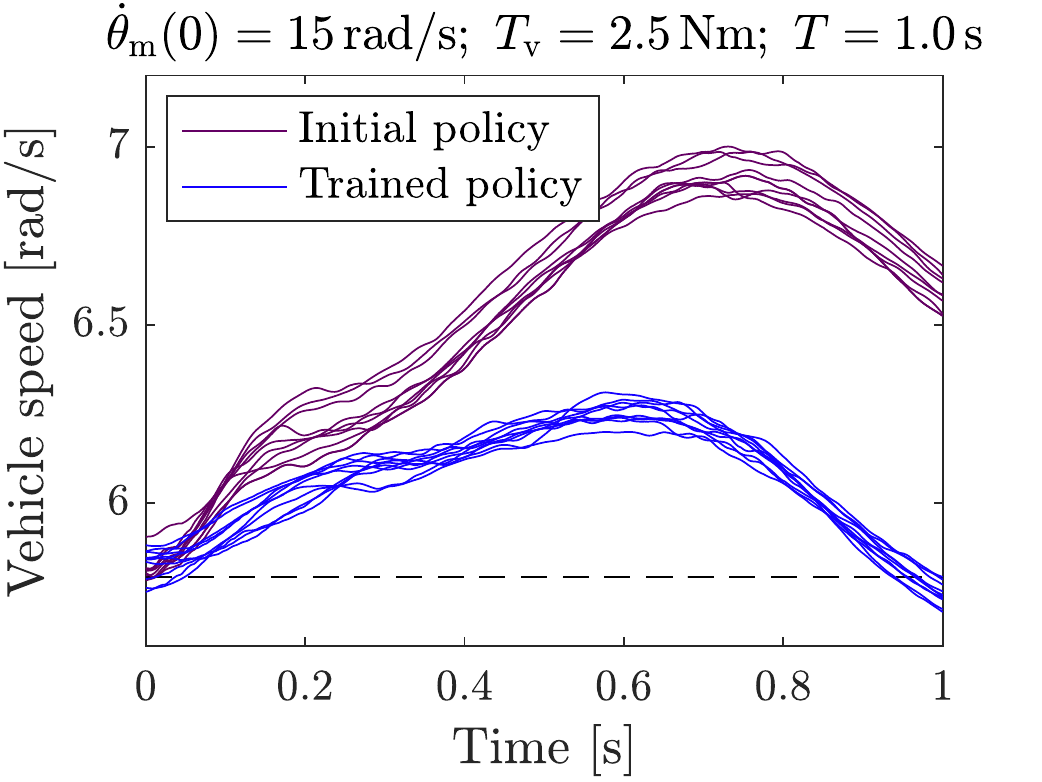}
		\caption{}
		\label{fig:shift_3}
\end{subfigure}
	\caption{Comparison of repeated gearshifts between the initial (purple) and trained (blue) policies (a) under the same operating conditions and gearshift trajectory used during the learning process, (b) under the same operating conditions, but with a shorter gearshift duration, and (c) at reduced motor speed and vehicle torque.}
	\label{fig:shift_situations}
\end{figure}

Moreover, Figure~\ref{fig:shift_situations} shows the repeatability of the results. Figure~\ref{fig:shift_1} shows 10 gearshift trials with the initialized policy (in purple), and 10 gearshift trials with the learned policy (in blue). This indicates that the improvement reported in Figure~\ref{fig:iterations} and Table~\ref{tbl:error} are not due to mere variations in the measurements. Figures~\ref{fig:shift_2} and \ref{fig:shift_3} show that the learned parameters also improve the gearshift quality for conditions that were never used during training. Figure~\ref{fig:shift_2} shows a gearshift with a shortened duration, i.e., $0.6 \, \mathrm{s}$ instead of the original $1.0 \, \mathrm{s}$. Figure~\ref{fig:shift_3} shows a $1\, \mathrm{s}$ gearshift initiated at reduced motor speed and reduced vehicle load. The practical implication this suggests is that the automatic tuning of gearshift controller parameters using the proposed method does not require trying a myriad of operating conditions, which greatly accelerates the tuning process. \\

\begin{figure}
	\begin{subfigure}[t]{.5\textwidth}
		\centering
		\includegraphics[width=0.75\linewidth,trim={0cm 0cm 0.5cm 0cm},clip]{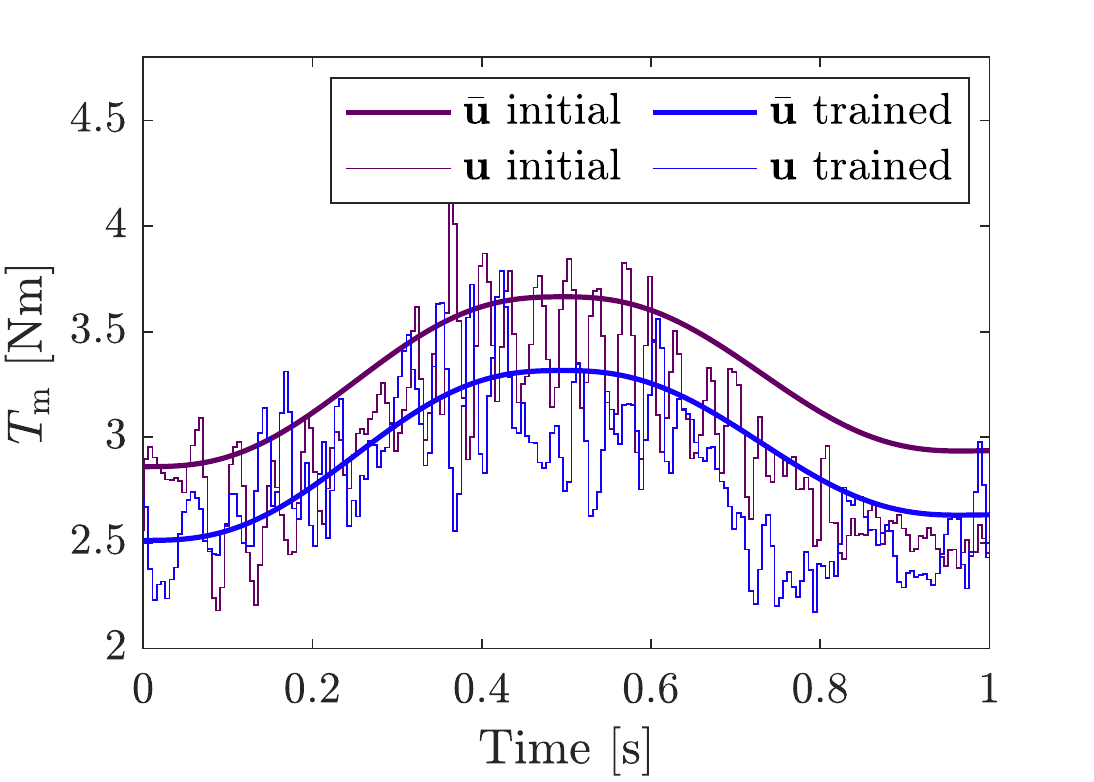}
		\caption{Motor torque}
		\label{fig:Tm}
	\end{subfigure}
	\begin{subfigure}[t]{.5\textwidth}
		\centering
		\includegraphics[width=0.75\linewidth,trim={0cm 0cm 0.5cm 0cm},clip]{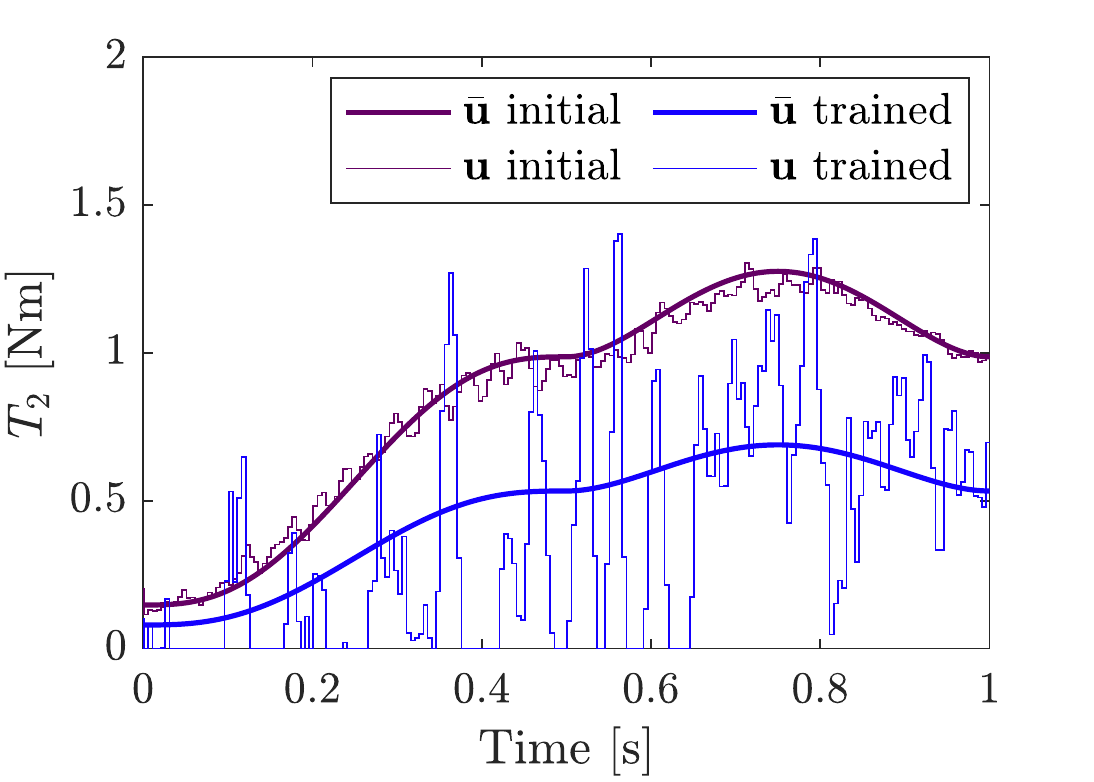}
		\caption{Clutch 2 torque}
		\label{fig:T2}
	\end{subfigure}
	\caption{Comparison of the control signals for the initial policy (purple) and the trained policy (blue). The thick lines represent the feedforward component of the control signals ($\bar{\bu}$), and the thin lines represent the total controller output ($\bu$), as in Equation~\eqref{eq:control} and Figure~\ref{fig:control}.}
	\label{fig:torques}
\end{figure}

Figure~\ref{fig:torques} shows how the learning process affects the torque commands. The two gearshift trials displayed -- initial policy and trained policy -- correspond to the trials with $i$ = 0 and $i$ = 4 in Figure~\ref{fig:iterations}, respectively. Figure~\ref{fig:Tm} shows that with the initial policy, the nominal motor torque ($\bar{\bu}$, thick purple line) is likely set too high, as the total controller output ($\bu$, thin purple line) is almost always lower than the nominal torque. The learning process reduces the nominal torque, which makes the controller's output more centered around the nominal value. This suggests that the learning method appropriately corrects feedforward parameters. Figure~\ref{fig:T2} shows that the feedback gains for the command of $T_2$ are greatly increased. The initial policy barely deviates from the nominal torque command, and the trained policy does so significantly. Note that the torque command is saturated at zero, since it is impossible to command a negative torque on a friction clutch. This large increase in the feedback gains, combined with the fact that it improves the trajectory tracking (see Figure~\ref{fig:iterations} and Table~\ref{tbl:error}), suggests that the feedback did not have enough authority in the initial policy, which the learning process corrects. It also belies unknown dynamics in the physical transmission, and motivates the learning approach.\\

Finally, Figure~\ref{fig:iterations} and Table~\ref{tbl:error} show that the policy seem to converge to a local minimum. In practice, engineers could restart the learning process with different initial values for the controller parameters, which would help determine whether the local minimum is also a global one given the current controller parametrization. Afterward, engineers could vary the parametrization of the controller, and determine whether it is possible to further improve the gearshift performance. After all, the chosen feedback controller and rather arbitrary parametrization of the feedforward signal may not be the best one can devise. This highlights the importance of a flexible approach that can learn quickly. Here, we argue the approach is very flexible due to the use of automatic gradients. And because the method can learn from few gearshift trials, several parametrizations can be tried. 

\section{Conclusion}
In this work, we calibrated the parameters of a gearshift controller with reinforcement learning. Our results show that the proposed method can tune numerous feedforward and feedback parameters concurrently, and it does so using only a few gearshift trials. The calibrated controller also performs better under operating conditions that are outside the training data set. The method can easily be adapted to various controller types due to the use of automatic differentiation. The proposed method should accelerate the calibration of gearshift controllers in a transmission development process, especially compared to methods inspired by design of experiments. The approach can be applied directly to other transmission architectures and control strategies, essentially covering the entire space of transmissions for electric vehicles. As a result, this should lead to better gearshift controllers, as more controller types and parametrization could be tried in a given project timeline. 

\section*{Funding sources}
This work was supported by Mitacs and Quebec's Fonds de recherche Nature et technologies.

\bibliographystyle{IEEEtranMdoi}
\bibliography{main}

\appendix
\section{Analytical gradients} \label{apx:grad}
This appendix provides analytical expressions for the computation of 
\begin{equation}
	\dod{J^\pi(\bpsi)}{\bpsi} = \sum_{t=0}^{T} \dod{}{\bpsi} \mathbb{E}_{\xt}[c(\xt)], \tag{\ref{eq:grad}}
\end{equation}
where $\bpsi$ is simply $\Kc$. Note that in this work, we use the numerator layout when displaying the Jacobian of a function. While a few of the results in this appendix can be found in the literature, see~\cite{mchutchon_nonlinear_2014} for instance, most are new results that pertain to our specific implementation of \textsc{pilco}.

\subsection{Gradients for the square exponential kernel function}
The square exponential kernel can be differentiated as follows:
\begin{align}
	\dpd{k(\bzs,\bzone)}{\bzs} &= -(\bzs-\bzone)^\top \Lambda^{-1} k(\bzs,\bzone), \\
	\dpd[2]{k(\bzs,\bzone)}{\bzs} &= -\Lambda^{-1}(I-(\bzs-\bzone)(\bzs-\bzone)^\top \Lambda^{-1}) k(\bzs,\bzone).
\end{align}
This allows to obtain derivatives of the mean and variance functions with respect to a deterministic test point $\bz_*$. 
\begin{align}
	\dpd{\mu_d(\bzs)}{\bzs} &= -(\Kzsz \odot \by_d^\top(\Kzz + \sigma_\epsilon^2 I)^{-1}) \tilde{Z}_*^\top \Lambda^{-1}, \tag{\ref{eq:delmud}}\\
	\dpd{\Sigma_d(\bzs)}{\bzs} &= 2(\Kzsz \odot \Kzsz (\Kzz + \sigma_\epsilon^2 I)^{-1}) \tilde{Z}_*^\top \Lambda^{-1},
\end{align}
where $\tilde{Z}_* = [\bz_*-\bzone, \ldots, \bz_*-\bzn]$, and $\odot$ represents an element-wise product. The second derivative of the mean function can also be obtained. The index notation is used since a third order tensor needs to be introduced. 
\begin{align}
	\dpd[2]{\mu_d(\bzs)}{\bzs} &= (\hat{Z}_*)_{ijk}(\Kzsz^\top \odot (\Kzz + \sigma_\epsilon^2 I)^{-1} \by_d )_k, \\
	\mathrm{where} \quad (\hat{Z}_*)_{ijk} &= -\Lambda^{-1}(I-(\bzs-\bz_k)(\bzs-\bz_k)^\top \Lambda^{-1}).
\end{align}

\subsection{Gradients for the state distribution}
As a reminder, the equations for the state transition are
\begin{align}
	\muxt &= (\Ad - \Bd \Kc)\muxtm + \muf(\muztm) + \Bd(\ubtm + \Kc \xbtm) + \tzero, \tag{\ref{eq:trmu}}\\
	\Sxt &= (\Ad - \Bd \Kc) \Sxtm (\Ad - \Bd \Kc)^\top + \Sigmaf(\ztm), \tag{\ref{eq:trs}}
\end{align}
where $\bx \in \mathbb{R}^D$ and $\bu \in \mathbb{R}^F$. \\

The first gradient from Equations~\eqref{eq:dmudpsi}-\eqref{eq:dsdpsi} reported is
\begin{equation}
	\dpd{{\muxt}}{{\muxtm}} = \Ad - \Bd \Kc + \dpd{\muf({\muztm})}{{\muztm}} \dpd{{\muztm}}{{\muxtm}},
\end{equation}
where
\begin{align}
	\dpd{\muf({\muztm})}{{\muztm}} &= \begin{bmatrix}
		\dpd{\mu_1({\muztm})}{{\muztm}} \\
		\vdots \\
		\dpd{\mu_D({\muztm})}{{\muztm}}
	\end{bmatrix}, \\[5pt]
	\dpd{{\muztm}}{{\muxtm}} &= \begin{bmatrix}
		I_{D \times D} \\
		-\Kc
	\end{bmatrix}.
\end{align}
\\

The next term in Equation~\eqref{eq:dmudpsi} is trivial:
\begin{equation}
	\dpd{{\muxt}}{{\Sxtm}}=0.
\end{equation}
\\

Next, there is
\begin{align}
	\dpd{{\muxt}}{\bpsi} &= (\Bd)_{ij} (\xbtm-\muxtm)^\top_k + \dpd{\muf({\muztm})}{{\muztm}} \dpd{{\muztm}}{{\bpsi}}, \\
	\dpd{{\muztm}}{{\bpsi}} &= 
	\begin{bmatrix}
		\bm{0}_{D\times D \times F} \\ (I_{F\times F})_{ij} (\xbtm-\muxtm)^\top_k
	\end{bmatrix}.
\end{align}
\\

Next, there is
\begin{align}
	\dpd{\Sxt}{{\muxtm}} &= \dpd{\Sigmaf(\ztm)}{{\muxtm}}, \\[5pt]
	\left( \dpd{\Sigmaf(\ztm)}{{\muxtm}} \right)_{iij} &= \left( \dpd{\Sigma_i(\ztm)}{{\muxtm}} \right)_{1j1}, \\[5pt]
	\dpd{\Sigma_d(\ztm)}{{\muxtm}} &=  \Bigg( \dpd{\Sigma_d(\muztm)}{{{\muztm}}} + \dpd{}{{{\muztm}}} \bigg( \underbrace{\dpd{\mu_d(\muztm)}{{\muztm}} \Sztm \dpd{\mu_d(\muztm)}{{\muztm}}^\top}_\square \bigg)  \Bigg) \dpd{{\muztm}}{{\muxtm}}, \\[2pt]
	\dpd{(\square)}{{\muztm}} & = 2 \dpd{\mu_d(\muztm)}{{\muztm}}\Sztm \dpd[2]{\mu_d(\muztm)}{{\muztm}}.
\end{align}
\\

Next, there is
\begin{align}
	\left( \dpd{{\Sxt}}{{\Sxtm}} \right)_{ijkl} &= (\Ad - \Bd \Kc)_{ik} (\Ad - \Bd \Kc)_{jl}  + \dpd{\Sigmaf(\ztm)}{{\Sxtm}}, \\
	\left( \dpd{\Sigmaf(\ztm)}{{\Sxtm}} \right)_{iijk} &= \left( \dpd{\Sigma_i(\ztm)}{{\Sxtm}} \right)_{1jk1}, \\[2pt]
	\dpd{\Sigma_d(\ztm)}{{\Sxtm}} &= \dpd{(\square)}{{\Sztm}} \dpd{{\Sztm}}{{\Sxtm}}, \\[2pt]
	\dpd{(\square)}{{\Sztm}} &= \left(\dpd{\mu_d(\muztm)}{{\muztm}}\right)^\top \left(\dpd{\mu_d(\muztm)}{{\muztm}}\right), \\[2pt]
	\dpd{{\Sztm}}{{\Sxtm}}& = 
	\begin{bmatrix}
		(I_{D \times D})_{ik}(I_{D \times D})_{jl} & (I_{D \times D})_{ik}(\Kc)_{jl} \\
		(\Kc)_{ik}(I_{D \times D})_{jl} & (\Kc)_{ik}(\Kc)_{jl}
	\end{bmatrix}.
\end{align}
\\

Finally, there is
\begin{align}
	\dpd{{\Sxt}}{\bpsi} &= \dpd{}{\bpsi} \Big( \underbrace{(\Ad - \Bd \Kc) \Sxtm (\Ad - \Bd \Kc)^\top}_\lozenge \Big) + \dpd{\Sigmaf(\ztm)}{\bpsi}, \\[2pt]
	\dpd{(\lozenge)}{\bpsi} &= -\Big( \big((I_{D \times D})_{ik}(\Sxtm)_{lj}\big)_{imkl}\big(\Ad - \Bd \Kc\big)_{jm} \nonumber \\
	& + \big((\Ad - \Bd \Kc) \Sxtm\big)_{im} \big((I_{D \times D})_{jk}(I_{D \times D})_{il}\big)_{mjkl} \Big)_{ijkl} \Big( (\Bd)_{ik} (I_{D \times D})_{jl} \Big)_{kl}, \\[2pt]
	\dpd{\Sigma_d(\ztm)}{\bpsi} &= \dpd{\Sigma_d(\muztm)}{\bpsi} + \dpd{(\square)}{\bpsi},\\[2pt]
	\dpd{\Sigma_d(\muztm)}{\bpsi} &= \dpd{\Sigma_d(\muztm)}{{\muztm}} \dpd{{\muztm}}{\bpsi}, \\[2pt]
	\dpd{(\square)}{\bpsi} &= \dpd{(\square)}{{\muztm}} \dpd{{\muztm}}{\bpsi} + \dpd{(\square)}{{\Sztm}} \dpd{{\Sztm}}{\bpsi}, \\[2pt]
	\dpd{{\Sztm}}{\bpsi} &= \begin{bmatrix}
		\bm{0}_{D \times D \times F \times D} & (I_{F \times F})_{jk} (\Sxtm)_{il}\\ 
		(I_{F \times F})_{ik} (\Sxtm)_{lj} & \triangle
	\end{bmatrix}, \\[2pt]
	\triangle &= \big( (I_{F \times F})_{ik} (\Sxtm)_{lj} \big)_{imkl} \big( \Kc \big)_{jm} + \big( \Kc \Sxtm \big)_{im} \big( (I_{F \times F})_{jk} (I_{D \times D})_{il} \big)_{mjkl}.
\end{align}

\subsection{Gradients for the cost function}
The derivatives of the expected cost with respect to the state distribution are
\begin{align}
	\dpd{\mathbb{E}_{\xt}[c(\xt)]}{{\muxt}} &= \big(1-\mathbb{E}_{\xt}[c(\xt)]\big) (\muxt - \xbt)^\top \tilde{S}, \\[2pt]
	\dpd{\mathbb{E}_{\xt}[c(\xt)]}{{\Sxt}} &= \tfrac{1}{2} \big( 1- \mathbb{E}_{\xt}[c(\xt)] \big ) \Big( \tilde{S} - \big((\muxt - \xbt)(\muxt - \xbt)^\top\big)_{kl}\big(\tilde{S}_{ik} \tilde{S}_{jl}\big)_{lkij}  \Big).
\end{align}

\end{document}